\documentclass{article}

\usepackage{PRIMEarxiv}

\usepackage[utf8]{inputenc} % allow utf-8 input
\usepackage[T1]{fontenc}    % use 8-bit T1 fonts
\usepackage{hyperref}       % hyperlinks
\usepackage{url}            % simple URL typesetting
\usepackage{booktabs}       % professional-quality tables
\usepackage{amsfonts}       % blackboard math symbols
\usepackage{nicefrac}       % compact symbols for 1/2, etc.
\usepackage{microtype}      % microtypography
\usepackage{lipsum}
\usepackage{fancyhdr}       % header
\usepackage{graphicx}       % graphics
\graphicspath{{media/}}     % organize your images and other figures under media/ folder
\graphicspath{ {./images/} }
\usepackage{subcaption}
\title{Analysis of Nepotism in Bollywood using Personalized PageRank and Effective Influence
} 
\author{
  Apoorv Jain \\
  MTech \\
  Indian Institute of Technology Delhi \\
  Delhi\\
  \texttt{eet232190@ee.iitd.ac.in} \\
}

\begin{document}
\maketitle

\begin{abstract}
Bollywood is one of the largest film-producing industries with a large worldwide audience. In this paper, we will try to find the most important stars in the era of 1990 to 2014, as well as try to use social network analysis methods and metrics to analyze the role of blood connections in getting opportunities in the industry. We created the actor relationship data of around 1000 debutants using OpenAI API and used a novel approach "Effective Influence" to study the effect of having a blood-related established actor inside the industry. We found that on an average every actor/director or actor/actor pair is reachable by a path of length at most 4 and a correlation of 0.6 indicating the advantage of having a blood connection inside the network in getting a good co-cast in the debut film. 
 
\end{abstract}

% keywords can be removed
\keywords{Social Network Analysis \and PageRank \and Bollywood Analysis}

\section{Introduction}
Bollywood is one of the biggest film industries in the world, renowned for its cinematic
diversity, and a unique perspective of the Indian culture. India is one of the biggest film factories in the world, competing with China and the US, and Bollywood produces the major portion of Hindi films. With a target audience of 1.2 Billion only in India, it generates around 110 billion rupees every year. The recent mega success of films like Jawan, Pathan and Dangal has shown the worldwide popularity of Bollywood films and film stars.

Every workplace is affected by the social dynamics of the workers [8] impacting the work opportunities of each other. Historical successes of stars show trends where personal ties shaped the career growth of actors. The depth of these relationships often extends beyond the screen, influencing the public's perception and industry dynamic, which won't be possible to include in this study. So we will try to use the past film success between individuals to make a graph network and find the influence of a node on the directors looking to cast new actors in their upcoming films.

The paper is organized into 6 sections, Introduction, Nepotism, Literature Survey, Methodology, Results and Conclusion. The methodology is composed of the details of the dataset and techniques used in the analysis. 
\section{Nepotism: A Special Favouritism}
\label{sec:headings}

Favouritism is in every industry where one favours someone because of a similar background like the same birth region, race, religion or some kind of interaction outside the workplace[10]. This makes one more likely to get the opportunity [7] apart from his talent or skills for the particular work, It becomes a bigger problem when one starts favouring others just because they are related to him or her by blood, a popular term for this is Nepotism which has been a widely talked topic in the last decade. This can lead to less deserving and talented people getting better roles in the industry [9], degrading the overall quality of the film. If the people working in the flagship films of the industry are not the best actors, then the films are bound to have lesser quality, and lesser worldwide impact.

It is a fact that India has never won an Oscar for an Original Hindi Film, even after more than 100 years of its inception, and being one of the biggest annual producers of films in the world, India has won only 10 Oscars till now with 20 nominations, out of which Indian(1983) and Gandhi(2009) were not even directed by an Indian. One ray of hope is that India won Best Original Song for Nattu Nattu song last year for RRR. But overall it can be stated as a very poor performance of Indian cinema on the global stage.

India has the largest population of more than 1.4 Billion in the world, and just surpassed China in 2023. With this huge population, India is still not able to generate quality world recognized films, then there must be a big problem in the system.

\section{Literature Survey}

Most of the related work that we found was based on the prediction of revenue of actors and box office collection of films. Sabdick Roy Chowdhury [1] has used social network analysis on the IMDB dataset to find the future trends of revenues of different stakeholders of Bollywood. There has been a study on the importance of different stars in Bollywood using centrality measures like degree, closeness and betweenness centrality. Lyric Doshi[2] uses the Hollywood Stock Exchange(HSX) to build a dynamic prediction of the box office collection of films. They use movie rating and SNA metrics like blog betweenness to represent the general buzz on the movie from the web and from bloggers. They gather posts from IMDb forums to generate sentiment metrics for positivity and negativity based on the discussion in the forums.

There are also studies based on the interaction of the stakeholders and individuals involved in the film business. Seung-Bo Park [3] has proposed character-net to generate the relationship graph of characters and dialogues to extract accurate movie stories such as classification of major, minor or extra roles, community clustering, and sequences via clustering communities composed of characters. Ilia Karpov [4] studied the movie industry community structure to highlight the role of the casting director in movie success. They showed that using additional knowledge in the "actor"-"casting director"-"talent agent"-"director" communication graph leads to better movie rating prediction. Wladston Viana [13] studied the impact of formation of the film's team including directors, producers and writers, they used topological and non-topological metrics to find correlation between these metrics and success parameters such as rating and box office collection.
\section{Methodology}
\subsection{Dataset}
We used the TIMDB dataset [11] collected by Rakesh Jain and Tushar Aneja using the TMDB API for Bollywood films in the period of 1960 to 2019 containing 4329 films. We will be using data from 1990 to 2014 for making the graph network and 2015 to 2019 data for analyzing insider and outsider actors' debut films.
The dataset contains all the relevant details like movie title, release year, actors, directors, IMDb rating and IMDb votes, used in our analysis.

Collecting Nepotism data in a particular timeline was a bigger challenge as there was no publicly available curated dataset containing relations of Bollywood stars. So we resorted to Open AI API for creating this dataset. We gave a prompt with a list of debutants in the timeline, and asked it to return a JSON formatted file with the star name, related star and relationship. There were still many problems with the output in terms of the inconsistency of the output format of all prompts because there is a limit of around 4000 tokens, we have to split the list of names in many prompts. The output has to be preprocessed to generate a consistent and complete dataset for nepotism. We found a total of 1000 debutants from 2015 to 2019, out of which 40 were insiders.

Prompt - “You are a Bollywood Relations Expert, i will give you a list of
names of bollywood stars, you need to tell me if they are related to
someone in the bollywood(by blood) or not by giving 1 or 0 as well as
name of the related star. Return the Output for each star in JSON format
like this- \{"star name"
:
" "
,
"is related"
:1 or 0,
"related star name"
:
" " \}
Please make the JSON compatible."

\subsection{Prepocessing}
The film dataset had missing values in the IMDB votes. We imputed the IMDB votes with the median value of non-missing IMDB votes, as the IMDB votes distribution was skewed. The rows with the missing actor's list were dropped. We normalized the IMDB votes with the Min-Max Scaler method, to scale it between 0 and 1.
To maintain consistency, we enforced a limit of a maximum of 10 actors in each film. One of the outliers found was 3 Idiots as it had a very large IMDB votes value as compared to other films.

\subsection{Difference between Insider and Outsider}
Before proceeding to the analysis, we need to define the definition of insider and outsider. So in this study, we will be using the term "Insider" for a person who had a related person already inside the industry at the time of debut or before 2015 and the relation between them is by blood, meaning children or siblings.
The "Outsider" is someone with no significant relation in the industry at the time of debut or before 2015.

% See awesome Table~\ref{tab:table}.

% \begin{table}
%  \caption{Sample table title}
%   \centering
%   \begin{tabular}{lll}
%     \toprule
%     \multicolumn{2}{c}{Part}                   \\
%     \cmidrule(r){1-2}
%     Name     & Description     & Size ($\mu$m) \\
%     \midrule
%     Dendrite & Input terminal  & $\sim$100     \\
%     Axon     & Output terminal & $\sim$10      \\
%     Soma     & Cell body       & up to $10^6$  \\
%     \bottomrule
%   \end{tabular}
%   \label{tab:table}
% \end{table}

\subsection{Graph Representation}
We will be representing Actors and Directors as nodes of the Graph network. The Edges between actor-actor or actor-director or director-director, are undirected and denote their collaboration in a common film in the past.
We will be using IMDB votes as the strength of this edge to approximate the success of their previous films, in which they worked together. We are using IMDB votes as a factor for the success of the film. Why normalized IMDB votes and not ratings?
The number of votes gives more insight into the
popularity of the film and the assumption that 
less rating isn’t gonna stop a movie from
earning money in India. The graph network created using this strategy consisted of 2388 nodes and 18679 edges.

\begin{figure}[h]
\centering
\includegraphics[width=0.7\textwidth]{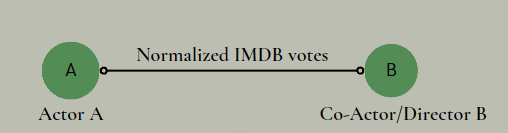}
\caption{Graph Representation}
\end{figure}

\subsection{Actor Importance}
For our further analysis, we need to find a measure of the importance of the stars in terms of their ability to attract the audience in the theatre, and in turn mega-hit films. We tried different centrality measures and Pagerank methods for this, 
\subsubsection{Degree centrality}
Degree centrality is the ranking where the node with a higher degree has a higher rank [5]. It can be crucial in finding the nodes with the most 1 hop reach useful in marketing applications. 
\[DegreeCentrality(v) = \frac{deg(v)}{max_{u \in  V}  deg(u)} \]
Degree centrality with value 1 is the node with the highest degree. Here Degree centrality gives the stars with the most number of films, which may not be so useful for our purpose as it doesn't include the fact that a lead role actor gains more popularity from a film than a supporting actor. Anupam Kher worked in highest films(244) mostly in supporting roles, and thus figure 2 shows supporting actors are more important than lead actors, that's why degree centrality is not the best metric for our purpose. 
\clearpage
\begin{figure}[h]
\centering
\includegraphics[scale=0.4]{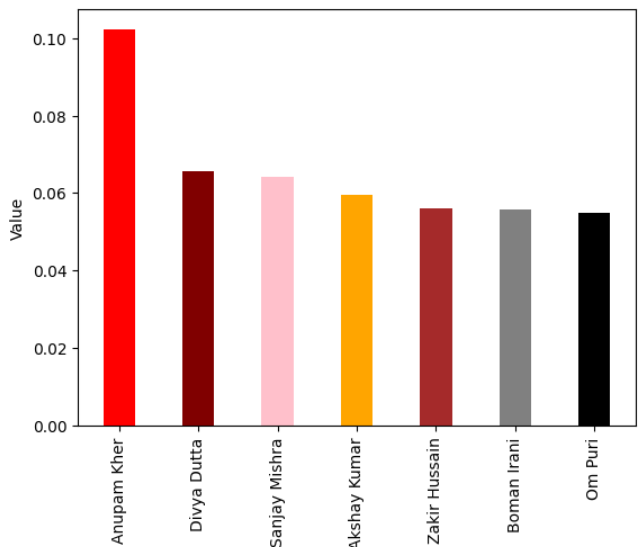}
\caption{Top 7 stars with Degree Centrality}
\end{figure}

\subsubsection{Closeness centrality}
Closeness centrality is the ranking where nodes which are closer to all other nodes have a higher rank [5]. These nodes can be used to spread information most efficiently throughout the network.
\[ClosenessCentrality(v) = \frac{(V -1)}{\sum_{u\epsilon V}d(u,v)}\]
where d(u,v) is the shortest path between node u and v.
Figure 3 shows the top 7 stars using closeness centrality, with a similar case as degree centrality.
\begin{figure}[h]
\centering
\includegraphics[scale=0.4]{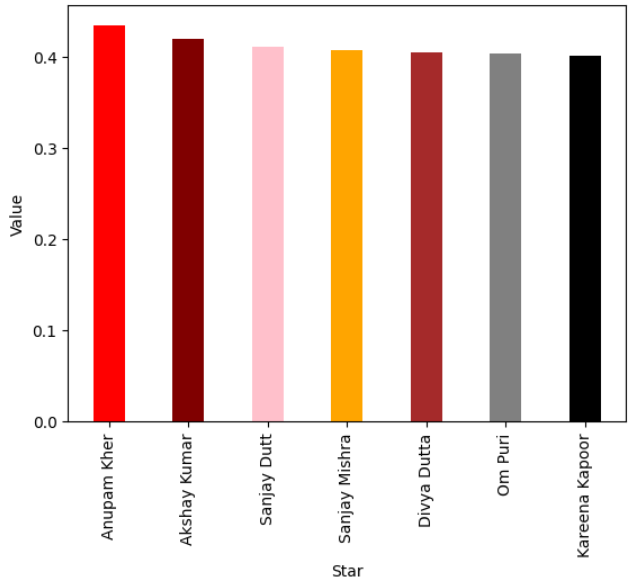}
\caption{Top 7 stars with Closeness Centrality}
\end{figure}
\subsubsection{Personalized Pagerank}
Pagerank was first proposed by Brin and Page[6] who used it on the World Wide Web, specifically for search engine results. The basic intuition of the algorithm was that important web pages will be pointed by important web pages. There was a dampening factor also introduced to take disconnected graphs into account in the random walk.

We used this algorithm on the actor-director graph with the intuition that the most important stars must have worked with other important stars in the past. One more important thing we added here is different weightage to each of the actors in a film, the lead actor will get the most popularity, and it will keep decreasing as the role number increases, considering the lead actor as 0.
Weightage of ith role in the jth film is given by, 
\[Weightage(i,j) = imdb(j)*\alpha ^{i}\]
where $\alpha$(importance attenuation factor) < 1 and imdb(j) gives the imdb votes of the film j.

This prevents the algorithm from giving a ranking with all the stars of the most popular film at the top, whether it is a lead actor or a supporting actor.

For Example,
Dhoom 3
['Aamir Khan
'
,
'Abhishek Bachchan
'
,
'Katrina Kaif'
,
'Uday Chopra
'
,
'Jackie Shroff']

Aamir Khan will gain $1/\alpha$ times more popularity from this film compared to Abhishek Bachchan.

The overall weightage of an actor should include all his films, and thus it is the average of all popularity gained in his films.
The top 7 stars using this are shown in figure 4, which shows the top 2 stars from lead actors of top hits like Dangal, PK and My Name is Khan by IMDB votes.

\begin{figure}[h]
\centering
\includegraphics[scale=0.4]{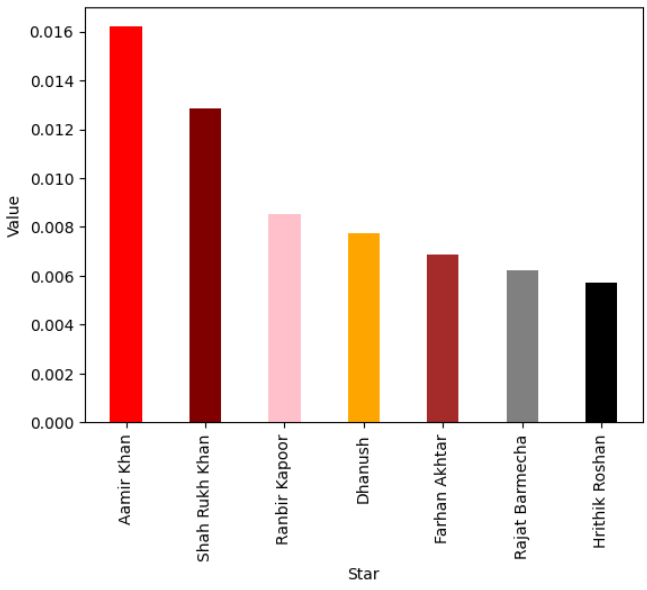}  
\caption{Top 7 stars with Personalized Pagerank}
\end{figure}

\subsection{Analysis of Insider vs Outsider Debut Films}

\subsubsection{Jaccard Coefficient}
Jaccard coefficient is a similarity index which uses the number of common neighbours to generate a similarity value between two nodes [5]. We will be using it to find the similarity between related actors and directors.
\[
J(u,v) = \frac{N(u) \cap  N(v)}{N(u) \cup N(v) )} 
\]
where u,v are two nodes and N(u) is the neighbourhood of u
\subsubsection{Adamic Adar Coefficient}
Adamic Adar is also a common neighbour strategy which predicts links in a social network, according to the number of common edges between two nodes [5]. It is defined as the sum of the inverse logarithmic degree centrality of the neighbours shared by the two nodes.
\[
A(u,v) = \frac{1}{\sum_{x\in N(u) \cap  N(v)}\left | N(x) \right |} 
\]
where u,v are two nodes and N(u) is the neighbourhood of u

\subsubsection{Effective Influence}
We have proposed Effective Influence to find the influence of a node on other nodes of the network as a function of importance and weights of links between intermediate nodes in the path between two nodes. There is an attenuation factor used to decrease the effect of the nodes as the path length increases because as the path length increases, the related actor's influence on the nodes also decreases.
The effective influence of Actor a on Director d is given by, 
\[Effective Influence(a,d) = PR(a) + \sum_{u \in Path(a,d)}^{} PR(u)*w(u)*(\alpha)^{|Path(a,u)|}\]
where PR(u) is the Pagerank importance of node u, Path(a,d) is the shortest path between node a  and node d, w(u) is the weight between u and the previous node in the path as described in figure 5 of path length 2 with an intermediate node I, and $\alpha$ is the attenuation factor with a value less than 1.

\begin{figure}[ht]
\centering
\includegraphics[scale=0.8]{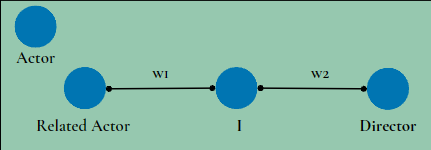}  
\caption{Effective influence on 2 length path}

\end{figure}
\section{Results}

Some of the interesting insights of this study, 1) The network follow real world properties i.e. power law, high clustering coefficient of 0.76 and small average path length of 3.2 . 2) The average path length of related actors with
top 100 directors are at most 4, which is inherently
present from the small world property. It establishes the reachability of an existing connection in the industry.(Refer Figure 6(b)) 3) There is a positive correlation of 0.5 and 0.6 implying that more the importance of the related actor and reachability, the greater the co-cast prominence in the debut film. (Refer to Figure 6 (a)) 4) The average rating of Insider debut movie is
about 1.0 rating less than outsiders which
maybe the reason for mediocre films.
5) The role number of insiders is found to be 2 less than
an outsider’s role number.
\begin{figure}[ht]
    \centering 
    \begin{subfigure}[b]{0.4\textwidth}
        \includegraphics[scale=0.4]{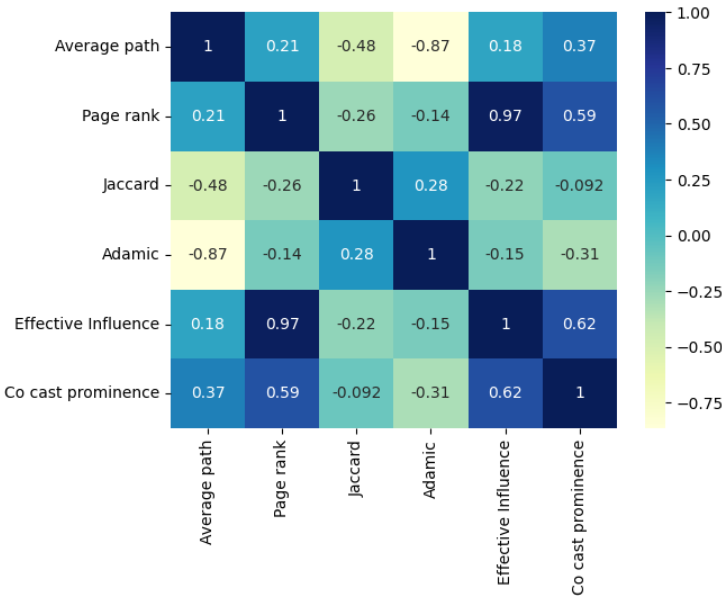}
        \caption{Correlation Matrix}
        \label{fig:subfig1}
    \end{subfigure}
    \hfill
    \begin{subfigure}[b]{0.4\textwidth}
        \includegraphics[scale=0.4]{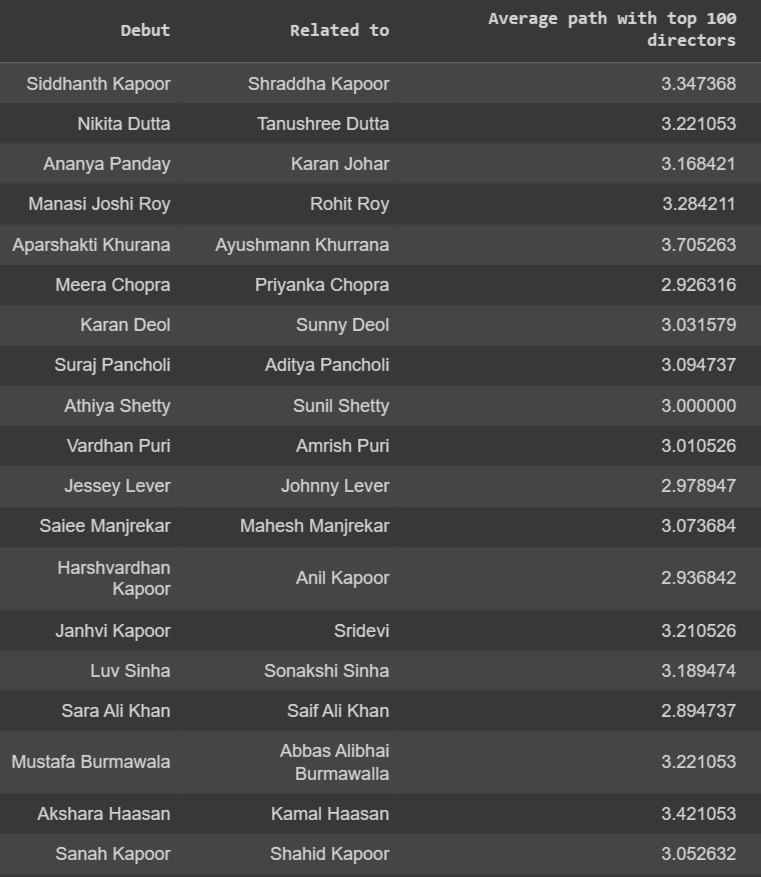}
        \caption{Average path length between related actor and top 100 directors}
        \label{fig:subfig2}
    \end{subfigure}
    
    \caption{Results}
    \label{fig:both}
\end{figure}

% \begin{figure}[h]

% \begin{subfigure}{0.5\textwidth}
% \includegraphics[scale=0.4]{Images/image.png}  
% \caption{Correlation matrix with co-cast prominence}   
% \end{subfigure}
% \begin{subfigure}{0.5\textwidth}
%     \includegraphics[scale=0.4]{Images/avgpath.png}  
%     \caption{Average path between related actors and top 100 directors}
% \end{subfigure}
    
% \end{figure}

\section{Conclusion}
 Like in every industry, Nepotism as a special favouritism play a role in the interaction of individual and thus impact the opportunities. We found a positive correlation of 0.6 using Effective Influence, indicating that the more the importance and reachability of the related actor, the more the chance of debutant getting into a film with a better cast. The network has a very small average path length of 4. We found the quality of films of insiders lower than outsiders. The number of outsiders debuting were more but they had lesser prominent roles in their debut films.

%Bibliography
\bibliographystyle{unsrt}  
\bibliography{references} 
\cite{sabdick_2016}
\cite{lyric_2010}
\cite{seung_2011}
\cite{ilia_2021}
\cite{sna_tan}
\cite{larry_1998}
\cite{allinthefamily_2018}
\cite{workplace_2022}
\cite{nepo_in_organization}
\cite{favouritism_work}
\cite{dataset}
\cite{box_office_pred}
\cite{movie_team_success}
\cite{hollywood_sna}
\end{document}